\documentclass[prl,aps,showpacs,epsf, twocolumn]{revtex4}
\usepackage{amssymb}
\usepackage{bm}
\usepackage[dvips]{graphicx}

\def\be{\begin{equation}}
\def\ee{\end{equation}}

\begin{document}

\title{Density profile of a trapped strongly interacting Fermi gas with
   unbalanced spin populations
}
\author{F. Chevy}
\affiliation{Laboratoire Kastler Brossel, \'Ecole normale
sup\'erieure, Paris, France }

\date{\today}

\begin{abstract}
We present a theoretical study of the density profile of a trapped
strongly interacting Fermi gas with unbalanced spin populations.
Making the assumption of the existence of a first order phase
transition between an unpolarized superfluid phase and a fully
polarized normal phase, we show good agreement with a recent
experiment presented in \cite{Partridge05}.
\end{abstract}


\maketitle

In the Bardeen-Cooper-Schrieffer (BCS) mechanism, the onset of
superfluidity is associated to pairing between two fermionic
species with matched Fermi levels. This scheme is relevant to
systems like  metal superconductors, superfluid $^3$He or ultra
cold gases, as observed in recent ground breaking experiments
\cite{Jochim03,Zwierlein03,Greiner03,Bourdel04,Kinast04,Partridge05b}.
However, other physical systems, like magnetized superconductors
or neutron stars, require the understanding of fermionic
superfluidity in the presence of mismatched chemical potentials.
The nature of superfluidity in these systems have been the subject
of longstanding debates \cite{Sarma63,Fulde64,Larkin65}, which
have been renewed by the  opportunity to address this topic
experimentally using gaseous samples. Various mechanisms were
proposed to describe the ground state of an ensemble of fermions
with mismatched Fermi levels: deformed Fermi surface
\cite{Sedrakian05}, FFLO (Fulde, Ferrell, Larkin, Ovshinnikov)
states, where Cooper pairs acquire finite momentum, and their
generalization to trapped systems
\cite{Combescot05,Mora05,Castorina05}, interior gap superfluidity
\cite{Liu03}, or phase separation between a normal and a
superfluid state through a first order phase transition
\cite{Bedaque03,Caldas05,Carlson05,Cohen05}. When the strength of
the interactions is varied, a complicated phase diagram mixing
several of these scenarios is expected
\cite{Pao05,Son05,Sheehy05}.

\begin{figure}
\includegraphics[width=\columnwidth]{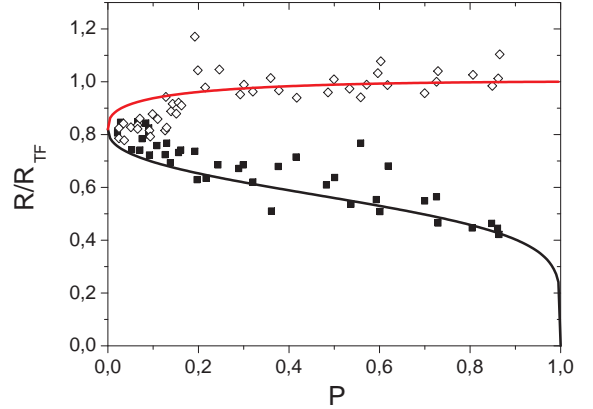} \caption{Comparison
between experimental data of \cite{Partridge05} and the model
presented here. $P=(N_1-N_2)/(N_1+N_2)$ is the imbalance between
the two spin populations. The radius $R$ of each component  is
normalized to the radius $R_{\rm TF}$ of the ideal Fermi gas
containing the same atom number. Squares and diamonds correspond
to experimental measurements of the radii of the minority and
majority components respectively. The full lines are the
predictions of the first order transition model.} \label{Fig1}
\end{figure}

Extending the seminal observation of fermionic superfluidity in
ultra-cold atom systems, two recent experiments
\cite{Zwierlein05,Partridge05} have started probing the regime of
mismatched Fermi levels by cooling samples containing different
atom number in each spin state. In this paper we will focus on the
results  presented in \cite{Partridge05} where the authors studied
the density profile of a gas of fermionic lithium when varying, in
the regime of strong interactions, the population imbalance
between the two trapped spin states. One the most striking results
is displayed in Fig. \ref{Fig1}. It shows that the radius of the
minority component is strongly reduced with respect to the one of
non-interacting gas with the same atom number. In the present
paper, we propose an interpretation of this result on the basis of
the existence of first order phase separation between the normal
and superfluid components and the use of universality in the
strong interaction regime. We show that these two ingredients are
sufficient to provide a good quantitative agreement with
experimental data. It is therefore complementary to recent studies
of \cite{Pieri05,Yi06,Silva06,Haque06}, who address the same
topics using a more complicated formalism.

Our analysis is based on the assumption of the existence of a zero
temperature first order phase transition  between a fully
polarized normal phase containing a single spin species and an
``unpolarized" superfluid state composed of a balanced mixtures of
the two species. The existence of this phase transition was
suggested by previous theoretical studies
\cite{Bedaque03,Caldas05,Carlson05} and is based on the following
qualitative argument: In the grand canonical ensemble, the
chemical potentials $\mu_i$ of the two species are fixed. In the
ground state, the system minimizes the grand-potential
$\Xi=H-\sum_i\mu_iN_i$, where $H$ is the hamiltonian of the system
and $N_i$ is the population of  species $i$. Let us now consider a
situation where the chemical potentials are mismatched, with
$\delta\mu=\mu_1-\mu_2>0$. Promoting a particle from state 2 to
state 1 decreases the grand potential  by $\delta\mu$ but implies
the breaking of a pair, hence increases the energy by the
superfluid gap $\Delta$. When $\delta\mu\lesssim\Delta$ we
therefore expect the system to be unpolarized, while above this
threshold a fully polarized system is obtained.

At zero temperature, the actual position of the phase transition
can be found analytically in the case of a homogeneous unitary gas
 by matching chemical potentials and pressures in
the two phases \cite{Carlson05}. Indeed, we can write in the
unpolarized phase $n_1=n_2=f(\mu_1,\mu_2)$. Using the
thermodynamical identity
$\partial_{\mu_1}n_2=\partial_{\mu_2}n_1$, we deduce that $f$ can
actually be expressed as a function of $\mu=(\mu_1+\mu_2)/2$ only.
To obtain the exact expression for $f$, we then consider the case
of matched Fermi surfaces, $\mu_1=\mu_2=\mu$. In this latter case,
universality at the unitary limit allows one to write $\mu=\xi
E_F$, where $E_F=\hbar^2(6\pi^2 n)^{2/3}/2m$ is the Fermi energy
of a non-interacting gas of density $n=n_{1,2}$ and $\xi$ is a
universal parameter whose determination has attracted a lot of
attention from both theoretical
\cite{Carlson03,Perali04,Astrakharchik04,Carlson05} and
experimental groups
\cite{Bourdel04,Bartenstein04,OHara02,Kinast05,Partridge05} and is
now thought to be $\xi\sim 0.45$. Using this expression for the
chemical potential, we then deduce that

\be
n_1=n_2=\frac{1}{6\pi^2}\left(\frac{m}{\xi\hbar^2}(\mu_1+\mu_2)\right)^{3/2},
\ee

Gibbs-Duhem identity $dP_S=n_1d\mu_1+n_2 d\mu_2$ finally yields
for the pressure in the superfluid phase

\be P_{\rm
S}=\frac{1}{15\pi^2}\left(\frac{m}{\xi\hbar^2}\right)^{3/2}\left(\mu_1+\mu_2\right)^{5/2}
\label{Eqn1}\ee

In the normal phase, we assume only the majority component is
present. We have then an ideal gas constituted of particles of
type 1 with chemical potential $\mu_1$, thus giving

\be P_{\rm
N}=\frac{1}{15\pi^2}\left(\frac{2m}{\hbar^2}\right)^{3/2}\mu_1^{5/2}.\label{Eqn2}\ee

Equating $P_{\rm N}$ and $P_{\rm S}$ \cite{SurfaceTension}, we see
that the two phases coexist only if $\mu_1$ and $\mu_2$ satisfy
the condition $\mu_2/\mu_1=\eta_{\rm c}$ with

\be \eta_{\rm c}=(2\xi)^{3/5}-1\sim -0.061,\label{Eqn3}\ee

\noindent a relation already found in \cite{Carlson05,Cohen05}. In
a trap, the chemical potential depends on position.

To compare with experiments, we now consider the case of a cloud
of atoms trapped in a harmonic potential $V(\bm r)=m\sum_i
\omega_i^2x_i^2/2$.
 Without loss of generality, we will
restrict our analysis to the case of a isotropic trap with
frequency $\bar\omega=(\omega_x\omega_y\omega_z)^{1/3}$. We can
indeed always recover the more general anisotropic case by making
the scaling transform $x_i\rightarrow \omega_i x_i/\bar\omega$.

To calculate the density profile of the cloud, we make use of the
local density approximation, where we assume that the chemical
potential of species $i$ depends on position as $\mu_i (\bm
r)=\mu_i^0-V(\bm r)$.

If we assume component 1 is the most populated, the inner
superfluid region is defined by the condition $\mu_2(\bm r)/\mu_1
(\bm r)<\eta_c$ and is bounded by the radius $R_2$ defined by

\be R_2^2=\frac{2}{m\bar\omega^2}\left(\frac{\mu_2^0-\eta_{\rm c
}\mu_1^0}{1-\eta_{\rm c}}\right). \ee

Atoms of the minority species are located in the paired superfluid
phase only. We thus have

\be N_2=\int_{r<R_2}n_2 (\bm r)\,d^3\bm
r=\frac{2}{3\pi\xi^{3/2}}\left(\frac{\mu_1^0+\mu_2^0}{\hbar\bar\omega}\right)^3g(R_2/\bar
R ), \label{Eqn4}\ee

\noindent where $\bar R^2=(\mu_1^0+\mu_2^0)/m\bar\omega^2$ and

\be g(x)=\frac{x\,{\sqrt{1 - x^2}}\,
     \left( -3 + 14\,x^2 - 8\,x^4 \right)  + 3\,\arcsin (x)}
    {48}.
\ee

Excess atoms of the majority species are located between $r=R_2$
and $r=R_1$ such that $m\bar\omega^2 R_1^2/2=\mu_1^0$. The number
of excess atoms is therefore $N_1-N_2=\int_{R_2}^{R_1} n_1 (\bm
r)\, d^3\bm r$, hence

\be
N_1-N_2=\frac{2}{3\pi}\left(\frac{2\mu_1^0}{\hbar\bar\omega}\right)^3(g(1)-g(R_2/R_1)).
\label{Eqn5}\ee

Dividing by (\ref{Eqn5}) by (\ref{Eqn4}) yields the implicit
equation for $\eta_0=\mu^0_2/\mu_1^0$ as a function of $N_1/N_2$

\be
\frac{N_1}{N_2}=1+\xi^{3/2}\frac{8}{(1+\eta_0)^3}\frac{g(1)-g(R_2/R_1)}{g(R_2/\bar
R)}. \label{Eqn6}\ee

\begin{figure}
\includegraphics[width=\columnwidth]{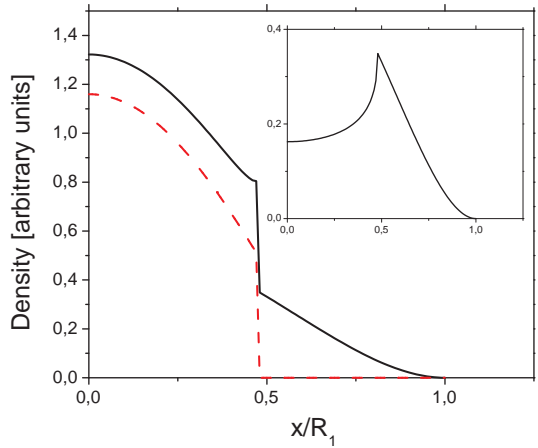} \caption{Integrated
column density of the majority (full line) and minority components
(dashed line) for a population imbalance $P=0.57$. Inset: excess
column density $\widetilde n_1-\widetilde n_2$. By contrast with
the experimental results, the boundary between the two phases is
here very sharp.} \label{Fig2}
\end{figure}

Equation (\ref{Eqn6}) is  solved numerically and the value
obtained for $\eta_0$ is then used to calculate the radii $R_1$
and $R_2$. The predicted evolution of the $R_i$ versus the
population imbalance $P=(N_1-N_2)/(N_1+N_2)$ is shown in Fig.
\ref{Fig1}. To follow Ref. \cite{Partridge05}, we have normalized
each $R_i$ to the Thomas-Fermi radius $R_{\rm TF}$ associated to
an ideal gas containing $N_i$ atoms. The agreement with the
experimental data is quite good as soon as $P\gtrsim 0.1$, a
remarkable result, since the model presented here contains no
adjustable parameter, as soon as the value of $\xi$ is known.

For weak population unbalance, experimental variations of the
radii is flatter than predicted by the first order transition
model. As proposed in \cite{Partridge05}, this suggests that the
phase separation does not happen exactly at $P=0$, but above some
threshold $P_{\rm c}\sim 0.1$. This point is strengthened when one
compare the theoretical and experimental density profiles. As in
\cite{Partridge05}, we have represented in Fig. \ref{Fig2} the
integrated column density $\widetilde n_i(x)=\int dy n_i (x,y,0)$
for $P=0.57$ (as in Fig. 2.D of \cite{Partridge05}). We
immediately see in this figure that the transition between the two
phases is very sharp, by contrast to what is observed
experimentally. Finite temperature might explain  this
discrepancy. Indeed, for low population imbalance, the superfluid
phase extends nearly throughout all the cloud, and in particular
in regions where the density, hence the Fermi energy, is very low.
In the unitary regime, the critical temperature for the
superfluid-normal transition is given by the scaling $k_{\rm
B}T_{c}=\alpha E_{\rm F}$, with $\alpha\sim 0.2$ \cite{Bulgac05}.
It may then happen that near the cloud edge, the temperature $T$
of the sample becomes locally larger than $T_{\rm c}$. If this
condition is satisfied at a radius $R_c$ smaller than the demixing
radius $R_2$, then the first order transition to the fully
unpaired state will not happen. If we introduce the Fermi
temperature $T_F=\hbar\bar\omega(6N_1)^{1/6}/k_{\rm B}$ associated
to a number of atoms $N_1$, the superfluid/normal and
paired/unpaired transitions will happen at the same position if
$T/T_{\rm F}$ satisfies the condition

\be \frac{T}{T_{\rm F
}}=\frac{\alpha}{2\xi}\left(\frac{R_1}{R_{\rm TF}}\right)^2
\left(1+\eta_0-2\left(\frac{R_2}{R_1}\right)^2\right).
\label{Eqn7}\ee

\begin{figure}
\includegraphics[width=\columnwidth]{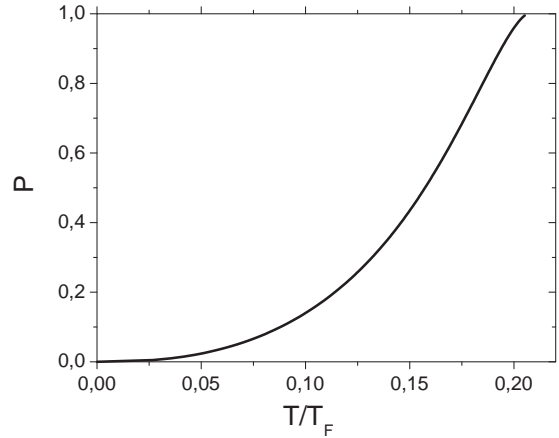} \caption{Critical population imbalance under which finite temperature effects overule demixing. The observed
experimental threshold $P\sim 0.1$ corresponds to a temperature
$T/T_{\rm F}\lesssim 0.1$} \label{Fig3}
\end{figure}

 In fig. \ref{Fig3}, we have plotted as a function of temperature the evolution of the critical imbalance under
which no demixing is expected. Experimentally,
 demixing only happens in the conditions of \cite{Partridge05} for $P\gtrsim 0.1$, which,
according to eq. \ref{Eqn7}, corresponds to $T/T_{\rm F}\lesssim
0.1$, a value compatible with experimental data. Despite a
semi-qualitative agreement, this finite temperature argument needs
to be clarified by a more careful analysis, following for instance
the work presented in \cite{Yi06}. Others scenarios can also be
envisioned to explain the smooth crossover between the two phases,
such as the existence of an intermediate phase -- e.g. a gapless
or FFLO phases as proposed in \cite{Son05} --, or a breakdown of
the local density approximation due to the fast variation of the
density profile. Nevertheless, the good agreement between theory
and experiment for the data presented in fig. \ref{Fig1} suggests
that the crossover region between paired and unpaired phases
should remain relatively narrow.

From the analysis presented here, it appears that the observations
of \cite{Partridge05} are consistent with a scenario of a
transition between a fully paired and a fully polarized phases
separated by a narrow crossover region probably related to finite
temperature effects breaking the superfluid phase before demixing
may occur. Interestingly, if this thermal scenario was confirmed
the measurement of the critical population unbalance under which
no phase separation happens could provide a useful tool for
fermion thermometry. Another issue not addressed here is related
to the superfluid character of the system when the population
imbalanced is varied. This is especially important if one wishes
to understand the results presented in \cite{Zwierlein05} where it
is shown that a mismatch of the Fermi surfaces by about 50\% leads
to a breakdown of superfluidity.

\acknowledgments The author wishes to thank C. Mora and C. Salomon
as well as the cold atom group for helpful discussions. This work
is partially supported by CNRS, Coll\`ege de France, ACI
nanoscience and R\'egion Ile de France (IFRAF). Laboratoire
Kastler Brossel is {\it Unit\'e de recherche de l'\'Ecole normale
sup\'erieure et de l'Universit\'e Pierre et Marie Curie,
associ\'ee au CNRS}.

\end{document}